\documentclass[3p,times]{elsarticle}

\usepackage{ecrc}

\usepackage{epstopdf}

\volume{00}

\firstpage{1}

\journalname{Nuclear Physics A}

\runauth{P.~Huovinen, P.~Petreczky, C.~Schmidt}

\jid{nupha}

\jnltitlelogo{Nuclear Physics A}

\usepackage{graphicx}
\usepackage{amsmath,amssymb}

\begin{document}

\begin{frontmatter}



\title{Equation of state at finite net-baryon density using Taylor
       coefficients up to sixth order}



\author[label1,label2]{Pasi~Huovinen}
\author[label3]{P\'eter~Petreczky}
\author[label4]{Christian~Schmidt}

\address[label1]{Institut f\"ur Theoretische Physik,
                 Johann Wolfgang Goethe-Universit\"at,
                 60438 Frankfurt am Main, Germany}
\address[label2]{Frankfurt Institute for Advanced Studies,
                 60438 Frankfurt am Main, Germany}
\address[label3]{Physics Department, Brookhaven National Laboratory,
                 Upton, NY 11973, USA}
\address[label4]{Fakult\"at f\"ur Physik, Universit\"at Bielefeld,
                 33615 Bielefeld, Germany}

\begin{abstract}
  We employ the lattice QCD data on Taylor expansion coefficients up
  to sixth order to construct an equation of state at finite
  net-baryon density. When we take into account how hadron masses
  depend on lattice spacing and quark mass, the coefficients evaluated
  using the p4 action are equal to those of hadron resonance gas at
  low temperature. Thus the parametrised equation of state can be
  smoothly connected to the hadron resonance gas equation of state. We
  see that the equation of state using Taylor coefficients up to
  second order is realistic only at low densities, and that at
  densities corresponding to $s/n_B \gtrsim 40$, the expansion
  converges by the sixth order term.
\end{abstract}

\begin{keyword}
lattice QCD \sep equation of state \sep hadron resonance gas

\end{keyword}

\end{frontmatter}



\section*{}

\begin{figure}
 \begin{center}
  \includegraphics*[width=140mm]{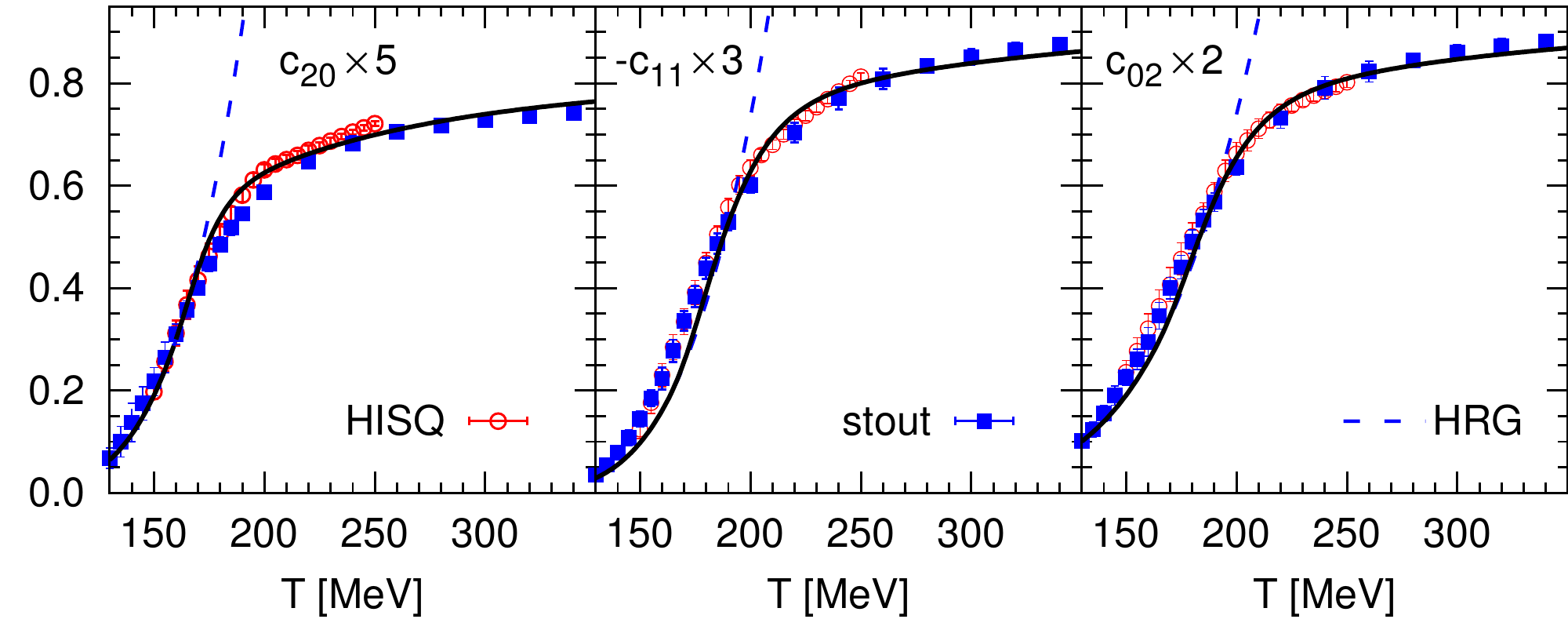}
  \caption{The parametrised (solid line) second order Fourier
    coefficients compared to HRG values (dashed) and the continuum
    extrapolated HISQ~\cite{Bazavov:2012jq} and
    stout~\cite{Borsanyi:2011sw} data.}
  \label{second}
 \end{center}
\end{figure}

One of the methods to extend the lattice QCD calculations to non-zero
chemical potential is Taylor expansion of pressure in chemical
potentials:
\begin{equation}
\frac{P}{T^4} = \sum_{ij} c_{ij}(T) \left(\frac{\mu_B}{T}\right)^i
                                  \left(\frac{\mu_S}{T}\right)^j.
 \label{PT4}
\end{equation}
The coefficients of this expansion are derivatives of
pressure, $P$, with respect to baryon and strangeness chemical 
potentials, $\mu_B$ and $\mu_S$, respectively:
\begin{equation}
 c_{ij}(T) = \frac{1}{i!j!}\frac{T^{i+j}}{T^4}\frac{\partial^i}{\partial \mu_B^i}
             \frac{\partial^j}{\partial \mu_S^j}P(T,\mu_B=0,\mu_S=0),
\end{equation}
where $T$ is temperature\footnote{We use natural units where
  $c=\hbar=k_\mathrm{B}=1$ throughout the text.}. The lattice QCD
calculations of these coefficients have matured to a level where both
Budapest-Wuppertal~\cite{Borsanyi:2011sw} and
hotQCD~\cite{Bazavov:2012jq} collaborations have published the final
continuum extrapolated results for the second order Taylor
coefficients, see Fig.~\ref{second}. 

\begin{figure}
 \begin{minipage}{0.47\textwidth}
  \begin{center}
   \includegraphics*[width=0.95\textwidth]{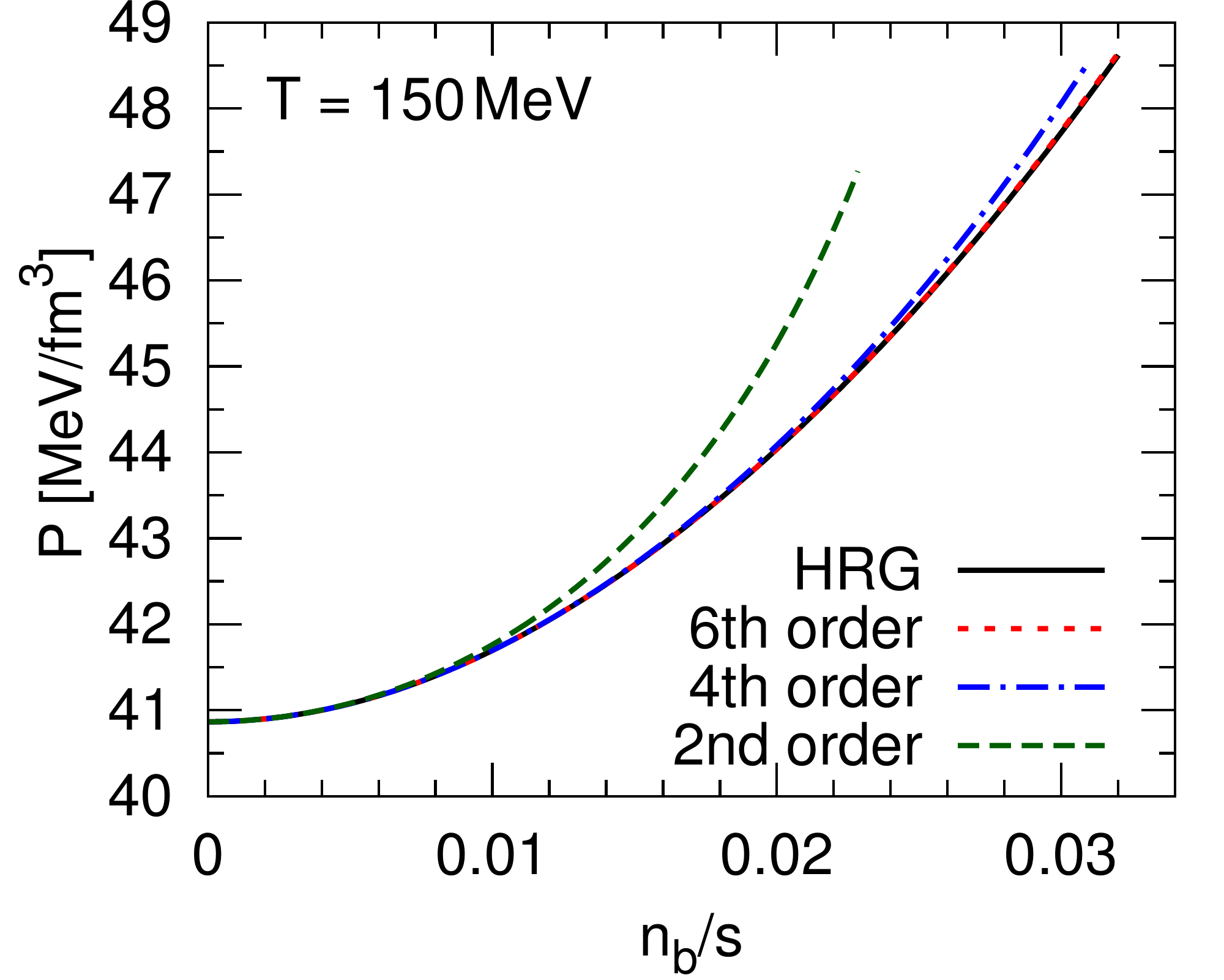}
   \caption{Pressure at constant temperature $T = 150$ MeV in hadron
     resonance gas (HRG) and using Taylor coefficients of HRG up to
     second, fourth and sixth order in Eq.~(\ref{PT4}).}
   \label{hrg}
  \end{center}
 \end{minipage}
  \hfill
 \begin{minipage}{0.47\textwidth}
  \begin{center}
   \includegraphics*[width=0.95\textwidth]{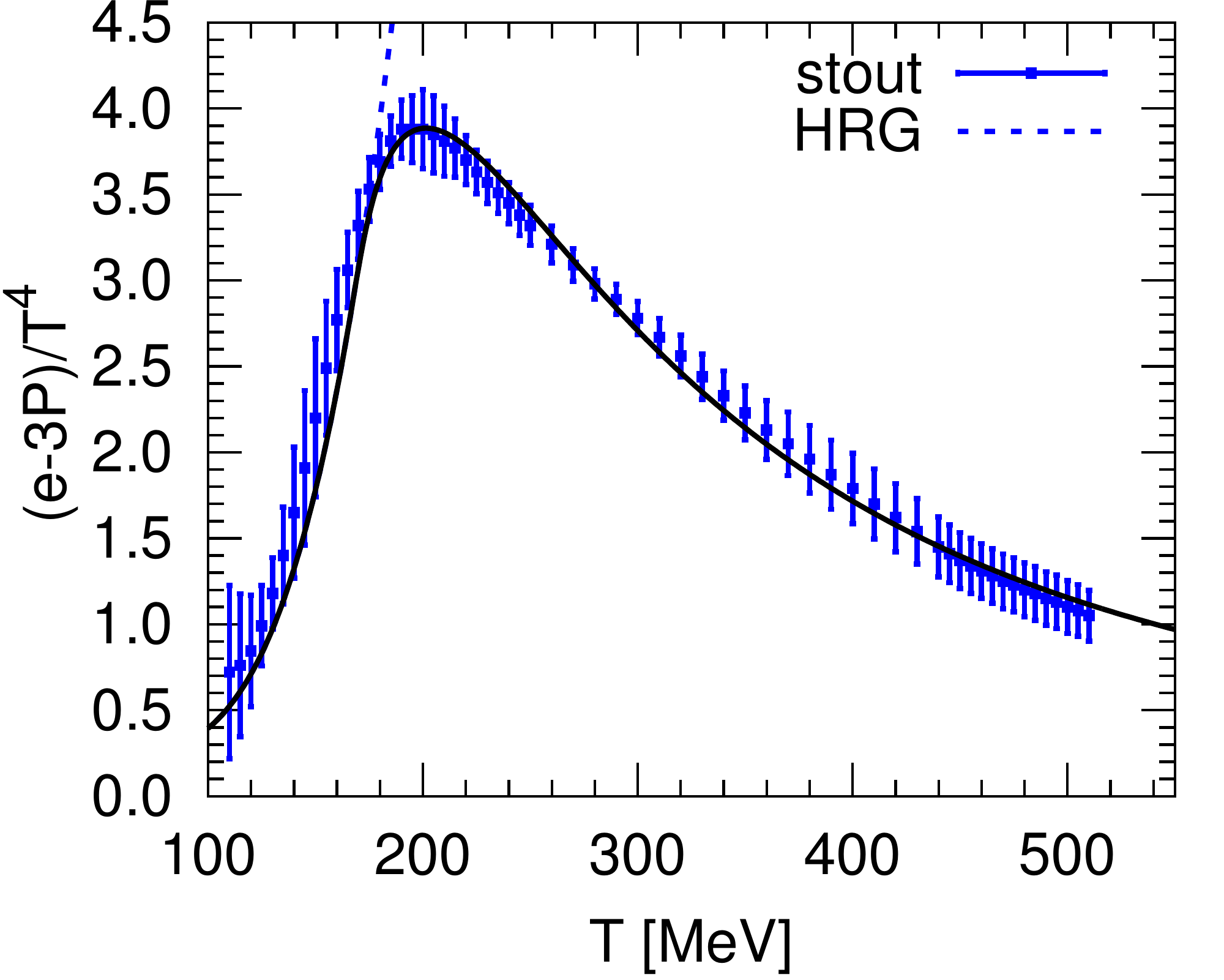}
   \caption{The fitted trace anomaly (solid curve) compared to the HRG
     trace anomaly (dotted) and the continuum extrapolated
     lattice QCD result using stout action~\cite{Borsanyi:2013bia}.}
   \label{trace}
  \end{center}
 \end{minipage}
\end{figure}

As seen in Fig.~\ref{second}, at low temperatures the coefficients
evaluated using the hadron resonance gas (HRG) model agree with the
lattice QCD results. Thus we may expect that HRG is an acceptable
approximation of the physical equation of state also at finite
net-baryon densities. To check how soon one may truncate the expansion
in Eq.~(\ref{PT4}), we calculate the Taylor coefficients in HRG up to
sixth order, evaluate the pressure using up to second, fourth or sixth
order Taylor coefficients, and compare to the actual HRG pressure. The
result is shown in Fig.~\ref{hrg}, where the pressure at constant 
$T = 150$ MeV temperature is shown as a function of inverse of entropy
per baryon, $n_B/s$. We use $n_B/s$ as variable to facilitate easy
comparison to heavy-ion collisions at various energies since, unlike
net-baryon density $n_B$, or baryon chemical potential $\mu_B$,
$s/n_B$ is (approximately) constant during the entire expansion stage
of the collision. We remind that at midrapidity the relevant entropy
per baryon is $s/n_B = 400$, 100, 65, and 40 at collision energies
$\sqrt{s_{NN}} \approx 200$, 64, 39, and 17 GeV/fm$^3$,
respectively. Thus an equation of state based on second order Taylor
coefficients only~\cite{Borsanyi:2012cr} can be expected to be
realistic only at relatively low net-baryon densities, $s/n_B \gtrsim
100$, \emph{i.e.}, at collisions with collision energy 
$\sqrt{s_{NN}} \gtrapprox 64$ GeV.

\begin{figure}
 \begin{center}
  \includegraphics*[width=\textwidth]{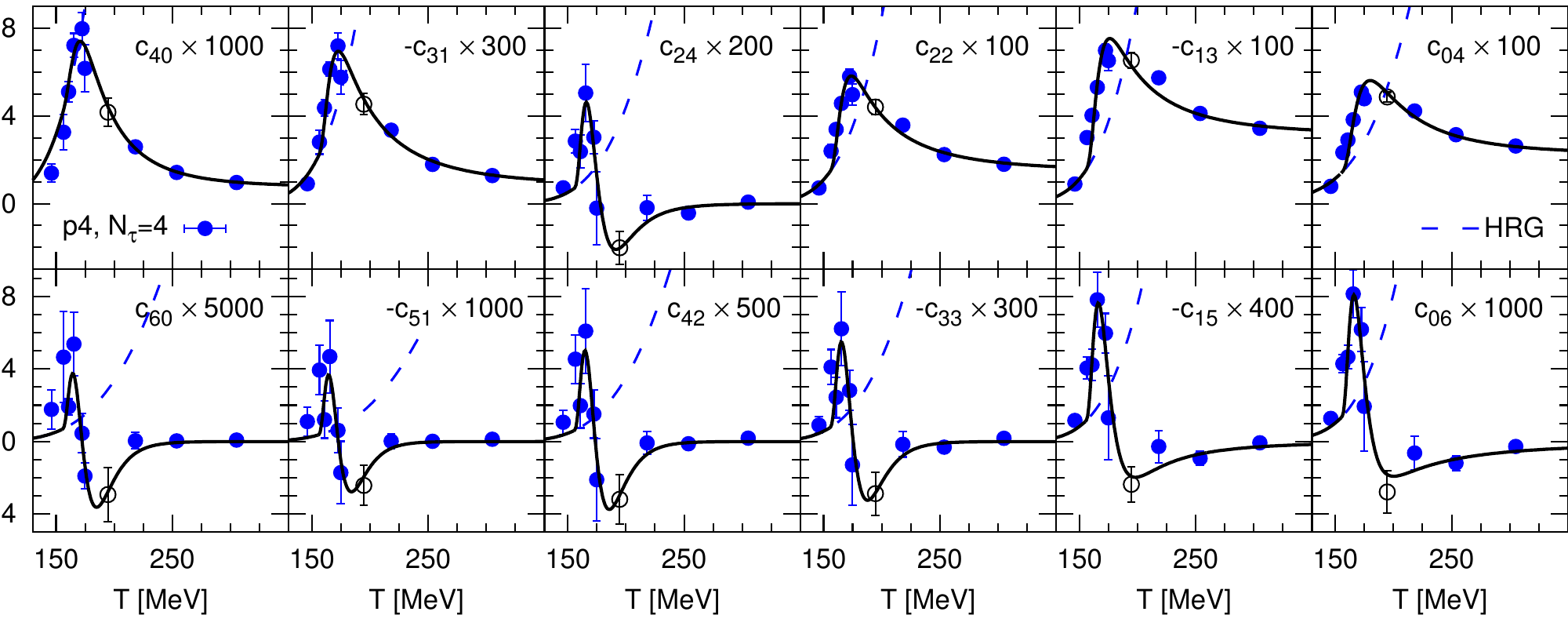}
  \caption{The parametrised (solid line) fourth and sixth order
    Fourier coefficients compared to HRG values (dashed line) and the
    shifted p4 data~\cite{Miao:2008,Cheng:2008} (see the text).}
  \label{stamps}
 \end{center}
\end{figure}

The Taylor coefficients have been evaluated on lattice up to sixth
order~\cite{Miao:2008,Cheng:2008}, but unfortunately the fourth and
sixth order coefficients suffer from large discretisation errors. As
we have argued
previously~\cite{Petreczky:2013qj,Huovinen:2009yb,Huovinen:2011xc,
  Huovinen:2012xm}, these errors are mostly due to the lattice
discretisation effects on hadron masses: When the hadron mass spectrum
is modified accordingly (for details see~\cite{Huovinen:2010tv}), the
HRG model reproduces the lattice data, see Fig.~1 of
Ref.~\cite{Huovinen:2011xc}. Interestingly this change can be
accounted for by shifting the modified HRG result of purely baryonic
coefficients towards lower temperature by 30
MeV~\cite{Huovinen:2011xc,Huovinen:2012xm}. Based on this finding, and
because the lattice data agree so well with the modified HRG, we
suggest that cutoff effects can be accounted for by shifting the p4
lattice data by 30 MeV. The fourth and sixth order coefficients are
shown in Fig.~\ref{stamps}, where the data below 206 MeV has been
shifted by 30 MeV, the data point at 209 MeV by 15 MeV (open symbol),
and the points above 209 MeV have not been shifted. At low
temperatures the shifted data now agrees with the unmodified HRG.

\begin{figure}
 \begin{center}
  \includegraphics*[width=125mm]{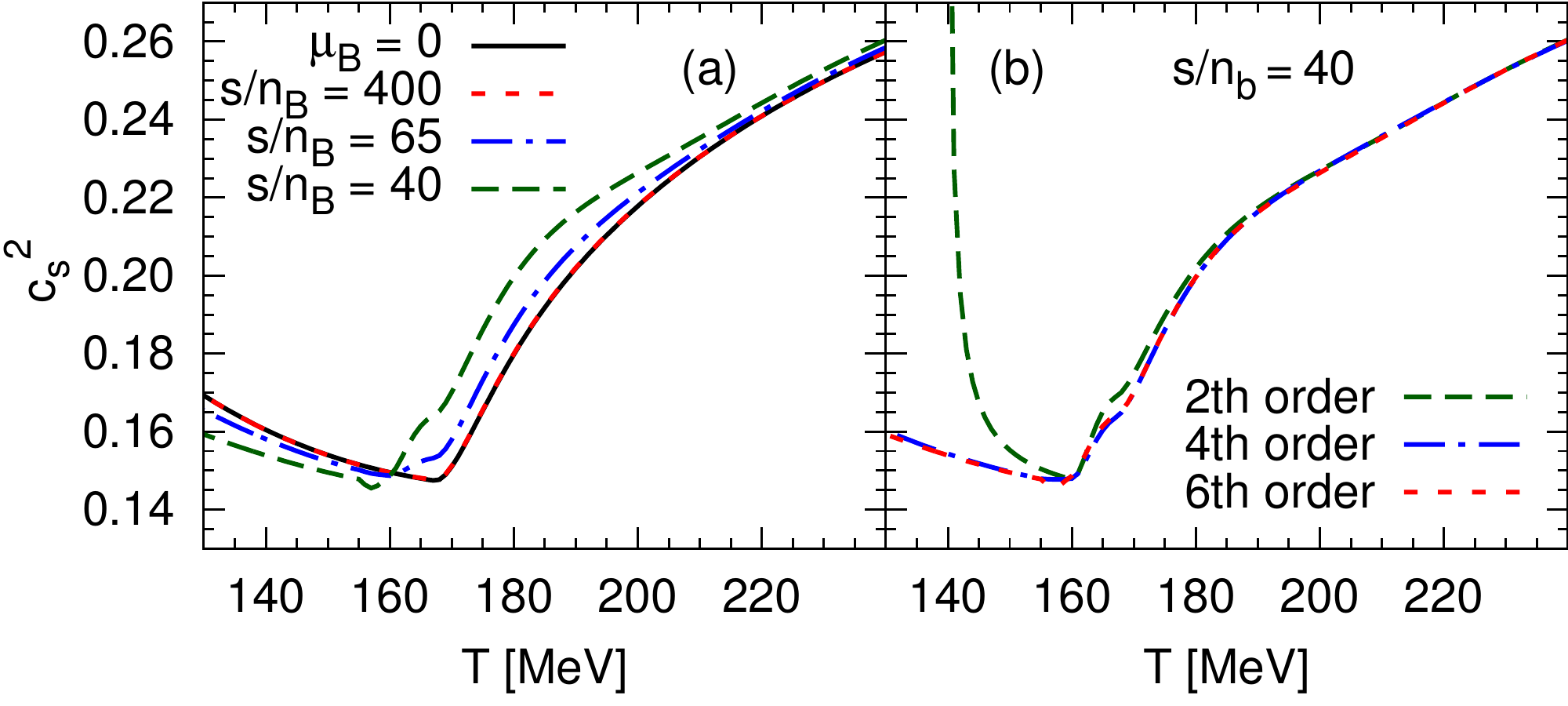}
  \caption{The square of the speed of sound, $c_s^2$, as a function of
    temperature on various isentropic curves with constant entropy per
    baryon (a), and on $s/n_B = 40$ curve evaluated using Fourier
    coefficients up to second, fourth or sixth order (b).}
  \label{sos}
 \end{center}
\end{figure}

We parametrise the shifted p4 data, and the unshifted, continuum
extrapolated stout and HISQ data, using an inverse polynomial of five
(second order) or six (fourth and sixth order coefficients) terms:
\begin{equation}
   c_{ij}(T) = \sum_{k=1}^m \frac{a_{kij}}{\hat{T}^{n_{kij}}} + c_{ij}^\mathrm{SB},
\end{equation}
where $c_{ij}^\mathrm{SB}$ is the Stefan-Boltzmann value of the
particular coefficient, $a_{kij}$ are the parameters, and the powers
$n_{kij}$ are required to be integers between 1 and 23. 
$\hat{T} = (T-T_s)/R$ with scaling factors $T_s=0$ and $R=0.15$ GeV
for the second order coefficients and $T_s=0.1$ GeV and $R=0.05$ GeV
for all other coefficients.  We match the parametrisation of second
order coefficients to the HRG value at temperature $T_\mathrm{SW}=160$
MeV, and the fourth and sixth order coefficients at
$T_\mathrm{SW}=155$ MeV by requiring that the Taylor coefficient and
its first, second, and third derivatives are continuous. The switching
temperatures have been chosen to optimise the fit and lead to smooth
behaviour of the speed of sound (see Fig.~\ref{sos}). These
constraints fix four (or five) of the parameters $a_{kij}$. The
remaining parameters are fixed by a $\chi^2$ fit to the lattice
data. The resulting parametrizations are shown as solid curves in
Figs.~\ref{second} and~\ref{stamps}.

We obtain the pressure at $\mu_B = 0$, \emph{i.e.} the coefficient
$c_{00}$, from the continuum extrapolated stout data for the trace
anomaly $(\epsilon - 3P)/T^4$~\cite{Borsanyi:2013bia}, which agree
with the very recent HISQ data~\cite{Bazavov:2014noa} within errors.
As in our earlier parametrisation of trace
anomaly~\cite{Huovinen:2009yb}, we fit the lattice result using an
inverse polynomial with four terms, and connect it to the HRG trace
anomaly at $T_\mathrm{SW} = 167$ MeV temperature, see Fig.~\ref{trace}.

We characterise the equation of state in Fig.~\ref{sos}a by showing
the square of the speed of sound, 
$c^2_s = \partial P/\partial\epsilon|_{s/n_B}$, on various isentropic
curves with constant entropy per baryon.  The curves at $s/n_B=400$,
65, and 40 are relevant at collision energies
$\sqrt{s_\mathrm{NN}}=200$, 39 and 17 GeV, respectively. At
$s/n_B=400$ (dotted line), the equation of state is basically
identical to the equation of state at $\mu_B = 0$ (thin solid line).
At larger baryon densities the effect of finite baryon density is no
longer negligible. The larger the density, the stiffer the equation of
state above, and softer below the transition temperature. We see some
ripples forming in the transition region with decreasing
$s/n_B$. These ripples grow fast with increasing density when one
goes beyond $s/n_B = 40$, and therefore we consider $s/n_B = 40$ to
give a practical maximum density for our parametrisation.

On Fig.~\ref{sos}b we have evaluated the square of the speed of sound
along the isentropic $s/n_B = 40$ curve using Taylor coefficients up
to second, fourth and sixth order. At temperatures above the
transition region they all lead to very similar speed of sound, but
below $T\approx 150$ MeV, the second order result deviates strongly
from the fourth and sixth order results. This shows that at low
temperatures equation of state based on the second order coefficients
only is not sufficient. Furthermore, the small difference between the
fourth and sixth order equation of state indicates that if $s/n_B
\gtrsim 40$, the expansion has basically converged by the sixth order
term.

To summarise we have argued that an equation of state based on the
Taylor expansion up to second order, is realistic only in collisions
with larger collision energy than at the RHIC beam energy scan. We
argue that the temperature shift of 30 MeV is a good approximation of
the discretisation effects in the lattice QCD data obtained using p4
action. We have constructed an equation of state for finite baryon
densities based on hadron resonance gas and lattice QCD data. In such
an equation of state, the Taylor expansion essentially converges by
the sixth order term if $s/n_B \gtrsim 40$.

\section*{Acknowledgements}
This work was supported by BMBF under contract no.\ 06FY9092, and by
the U.S. Department of Energy under contract DE-AC02-98CH1086.








\end{document}